\documentclass[aps,prb,twocolumn,groupedaddress,showpacs]{revtex4-1}%
\usepackage{epsfig}
\usepackage{amsmath}
\usepackage{amsfonts}
\usepackage{amssymb}
\usepackage{graphicx}%
\providecommand{\U}[1]{\protect\rule{.1in}{.1in}}
\begin{document}
\title{Symmetry analysis of possible superconducting states in K$_x$Fe$_2$Se$_2$ 
superconductors}
\author{I.I. Mazin}
\affiliation{Code 6393, Naval Research Laboratory, Washington, DC 20375, USA}

\begin{abstract}
A newly discovered family of the Fe-based superconductors 
is isostructural with the so-called 122 family of Fe pnictides,
but has a qualitatively different doping state. Early experiments 
indicate that superconductivity is nodeless, yet prerequisites for
the $s_\pm$ nodeless state (generally believed to be realized in Fe
superconductors) are missing.
It is tempting to assign a $d-$ wave symmetry to the new materials, 
and it does seem at first glance that such a state may be nodeless.
Yet a more careful analysis shows that it is not possible, given the particular
122 crystallography, and that the possible choice of admissible symmetries
is severly limited: it is either a conventional single-sign $s_{+}$
state, or another $s_\pm$ state, different from the one believed to be present
in other Fe-based superconductors.

\end{abstract}

\pacs{74.20.Pq,74.25.Jb,74.70.Xa}
\maketitle

Recent reports of superconductivity at $T_{c}$ in excess of 35
K\cite{firstreports} in Se-based iron superconductors (FeBS) isostructural
with BaFe$_{2}$As$_{2}$ (the so-called 122 structure) have triggered a new
surge of interest among the physics community. These materials are believed by
many to open a new page in Fe-based superconductivity. Indeed, the formal
composition, AFe$_{2}$Se$_{2},$ where A is an alkali metal, corresponds to a
formal doping of 0.5 electron off the standard for FeBS parent compounds
(LaFeAsO, BaFe$_{2}$As$_{2},$ or FeSe) valence state of iron, Fe$^{2+}$. Such
a large doping in other materials, such as Ba(Fe,Co)$_{2}$As$_{2}$ leads to a
complete suppression of supercondctivity, which has been generally
ascribed\cite{Wen,french} to disappearence of the hole pockets of the Fermi
surface and formal violation of the quasinesting condition for the $s_{\pm}$ superconductivity.

Indeed all band structure calculations show\cite{bands1} that in
AFe$_{2}$Se$_{2}$ the hole bands are well under the Fermi surface (for the
reported experimental crystal structure of KFe$_{2}$Se$_{2},$ about 60 meV),
and this is confirmed by preliminary ARPES
results\cite{ARPES1,ARPES2,DingTl}. This has led to speculations that in this
subfamily it is not the familiar $s_{\pm}$ superconductivity that is realized, but a
$d-$wave superconductivity\cite{ARPES1, Graser,Lee,Balat} of the sort
discussed in an early paper by Kuroki $et$ $al$\cite{Kuroki}. Unfortunately,
these speculations are entirely based upon the \textquotedblleft
unfolded\textquotedblright\ Brillouin zone description of the electronic
structure, a simplified model that neglects the symmetry lowering due to the
As or Se atoms and the fact that in the real unit cell there are two Fe ions,
and not one. Furthermore they implicitly assume that spin susceptibility
corresponding to the \textquotedblleft checkerboard\textquotedblright\ wave
vector, $Q=(\overline{\pi},\overline{\pi}),$ is substantially enhanced,
despite the fact that this vector corresponds to an electron-electron
interband transition that is much less efficient in enhancing susceptibility (here
and below, we used the bar when we work in the unfolded Brillouin zone). This
assumption is supported by model calculations based on an onsite Hubbard
Hamiltonian\cite{Graser}, but its applicability to FeBS is still an open question.

In this paper, we will critically address these two assumptrions, and will
show that the latter assumption is supported by first principles calculations,
but the former assumpion is actually very misleading. We will present a
general symmetry analysis of possible superconducting symmetries supported by
the Fermi surface topology existing in AFe$_{2}$Se$_{2}.$ This analysis is not
limited by a specific density functional calculation, but is based on the
general crystallographic considerations appropriate for this crystal
structure. It appears that it is impossible to fold down a nodeless $d-$wave
state so as to avoid formation of line nodes. Thus, emerging experimental
evidence from ARPES, \cite{ARPES1,ARPES2,DingTl}, specific heat\cite{WenSH},
NMR\cite{NMR}, and optics\cite{opt} that superconductivity in AFe$_{2}$%
Se$_{2}$ is nodeless is a strong argument against $d-$wave. A conventional
$s-$state is also unlikely based on the proximity to magnetism and actual
observation of a coexistance of superconductivity and magnetism. We emphasize
that the symmetry of the folded Fermi surfaces does allow for a nodeless state,
which however has an overall $s$ symmetry and can also be called $s_{\pm}$, as
it is strongly sign-changing. Unlike the $s_{\pm}$ advocated for the
\textquotedblleft old\textquotedblright\ FeBS it is not driven by
($\overline{\pi},0)$ spin fluctuations and cannot be derived from considering
an unfolded Brillouin zone Fermi surface.

\begin{figure}[t]
\includegraphics[width=0.8\columnwidth]{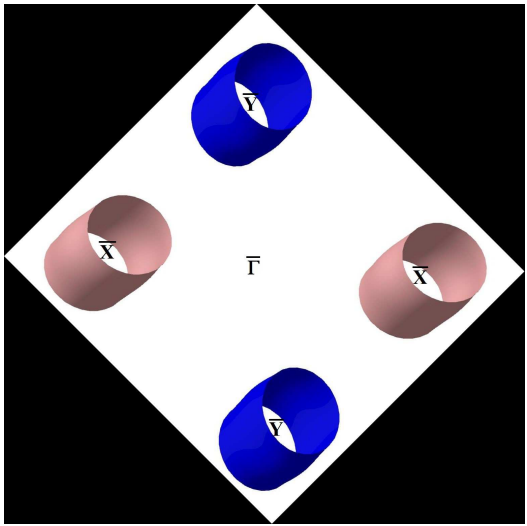}\caption{
A cartoon showing a generic 3D Fermi surface for an AFe$_2$Se$_2$
material in the unfolded (one Fe/cell) Brillouin zone. Different 
colors show the signs of the order parameter in a nodeless $d-$wave
state, allowed in the unfolded zone. The $\Gamma$ point is in the center
(no Fermi surface pockets around $\Gamma$), and the electron pockets are around the
$\bar{X}, \bar{Y}$ points.} 
\label{unfoldedFS}%
\end{figure}
The unfolded Fermi surface topology in materials with the 122 structure is
controlled by two factors: ellipticity of individual electron pockets and
their $k_{z}$ dispersion (Fig. \ref{unfoldedFS}). The ellipticity in the
unfolded zone is determined by the relative position of the $xy$ and $xz/yz$
levels of Fe, and the relative dispersion of the bands derived from them.
Indeed\cite{PALee}, the point on the Fermi surface located between
$\overline{\Gamma}$ and $\overline{X}$ has a purely $xy$ character, while that
between $\overline{\Gamma}$ and \={M} a pure $yz$ character. At the
$\overline{X}$ point the $xy$ state is slightly below the $yz$ state, but has
a stronger dispersion, therefore depending on the system parameters and the
Fermi level the corresponding point of the Fermi surface may be more removed
from $\overline{X},$ or less. In the 1111 compounds, the first to have been
investigated, the dispersion of the $xy$ band is not high enough to reverse the
natural trend, so the Fermi surface remains elongated in the $\overline
{\Gamma}\overline{X}$ (1,0) direction.

For both $xy$ and $xz/yz$ bands the hopping mainly proceeds via As (Se)
$p-$orbitals. The $xy$ states mainly hop through the $p_{z}$ orbital (see Ref.
\cite{Ole} for more detailed discussions), and $xz$ ($yz)$ via $p_{y}$
($p_{x})$ orbitals. If there is a considerable interlayer hopping between the
$p$ orbitals, whether direct (11 family) or assisted (122 family), the
ellipticity becomes $k_{z}-$dependent. For instance, in FeSe there is
noticeable overlap between the Se $p_{z}$ orbitals, so that they form a
dispersive band with the maximum at $k_{z}=0$ and the minimum at $k_{z}%
=\pi/c.$ Obviously, hybridization is stronger when the $p_{z}$ states are
higher, therefore the Fermi surface ellipticity is completely suppressed in
the $k_{z}$=0 plane, while rather strong in the $k_{z}=\pi/c$ plane, which
leads to formation of the characteristic \textquotedblleft
bellies\textquotedblright\ in the Fermi surface of FeSe. On the other hand,
$p_{x,y}$ orbitals in FeSe do not overlap in the neighboring layers, so the
$xz$ and $yz$ bands have very little $k_{z}$ dispersion, so that the inner
barrels of the electronic pockets in this compound are practically 2D.

In 122, the interlayer hopping proceeds mainly via the Ba (K) sites, and thus
the $k_{z}$ dispersion is comparable (but opposite in sign!) for the $xy$ and
$xz/yz$ bands. As a result, when going from the $k_{z}=0$ plane to the $k_{z}
=\pi/c$
plane the longer axis of the Fermi pocket shrinks, and the shorter expands, so
that the ellipticity actually changes sign.

Importantly, the symmetry operation that folds down the single-Fe Brillouin
zone when the unit cell is doubled according to the As (Se) site symmetry is
different in the 11 and 1111 structures, as compared to the 122 structure. In
the former case, the operation in question is the translation by $(\bar{\pi
},\bar{\pi},0),$ without any shift in the $k_{z}$ direction, in the latter by
$(\bar{\pi},\bar{\pi},\bar{\pi}).$ Thus the folded Fermi surface in 11 and in
1111 has full fourfold symmetry, while that in the 122 has such symmetry only
for one particular $k_{z,}$ namely $k_{z}=\pi/2c.$ Furthermore, in 122 the
folded bands are not degenerate along the MX (now the labels are without the
bars, that is, corresponding to the folded BZ), as they were in 11/1111. Finally,
there is a considerable (at least on the scale of the superconducting gap)
hybridization when the folded bands cross (except for $k_{z}=0).$

Now we are ready to analyze possible superconducting symmetries in the actual
AFe$_{2}$Se$_{2}$ materials. We shall not adhere strictly to the calculated
band structure and the Fermi surfaces, but rather consider several
possibilities allowed by symmetry. Let us start first from a $d-$wave state in
the unfolded BZ, as derived in Refs. \cite{Kuroki,Lee,Graser}. In Fig.
\ref{unfoldedFS} we show by the two colors the signs of the order parameter.
Obviously in the \textit{unfolded} BZ such a state has no nodes.

\begin{figure}[t]
\includegraphics[width=0.8\columnwidth]{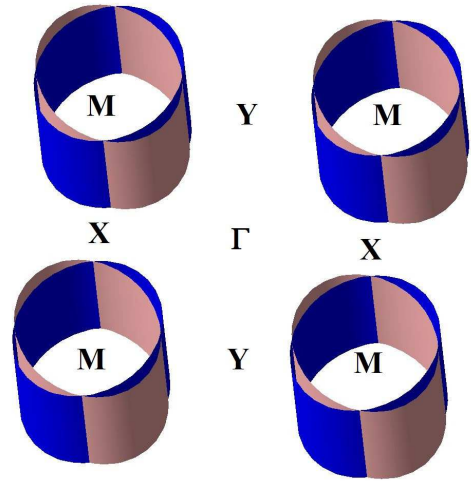}\caption{
A cartoon showing a folded 3D Fermi surface for an AFe$_2$Se$_2$
material, assuming a finite ellipticity, but zero $k_z$ dispersion.
  Different 
colors show the signs of the order parameter in a $d-$wave
state. Wherever the two colors meet, turning on hybridization
due to the Se potential creates nodes in the order parameter.} 
\label{dwave1}%
\end{figure}

Let us now assume that the $k_{z}$ dispersion is negligible, while the
ellipticity remains finite. After folding, but before turning on the
hybridization, we have the picture shown in Fig. \ref{dwave1}. The border between
the red and the blue colored regions now becomes a nodal line\cite{Parker}. In
this case, we have four such lines for each pair of electron pockets. One can
think of an effective \textquotedblleft thickness\textquotedblright\ of the
nodal lines, meaning the distance in the momentum space over which the sign of
the order parameter changes. This is defined by the ratio of the hybridization
gap at the point where the bands cross and their typical energy separation.
Analysis of the first principle calculations for both As and Se based 122
compounds indicates that this width is varying between zero (unless spin-orbit
interaction is taken into account) and a number of the order of 1. Thus, the
effect of the nodal lines on thermodynamical properties is comparable to that
in one-band $d-$wave superconductors such as cuprates and therefore should be
easily detectable.

\begin{figure}[t]
\includegraphics[width=0.8\columnwidth]{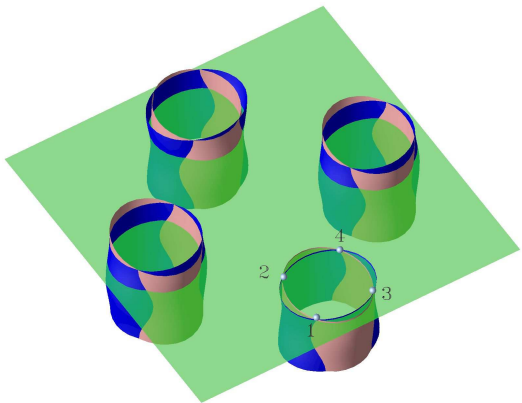}\caption{
Same as Fig. \protect\ref{dwave1}, but assuming a moderated
$k_z$ dispersion. The plane at $k_z=\pi/2c$ is shown, and one of the Fermi surfaces is clipped 
above this plane to show how the nodal points move away from their 
high symmetry positions.  } 
\label{3D}%
\end{figure}

Let us now gradually turn on the $k_{z}$ dispersion. Nothing changes for
$k_{z}=\pi/2c,$ that is, there are four equidistant nodes in this plane, which
we can label as 1, 2, 3 and 4. As we move towards $k_{z}=0,$ nodes 1 and 2 get
closer to each other, and so do nodes 3 and 4. As we move towards $k_{z}%
=\pi/c,$ the other pairs get closer, nodes 1 and 4, and nodes 2 and 3. Thus,
instead of four vertical node lines we get four wiggly lines, otherwise
similar in properties to the pure 2D case in Fig. \ref{3D}. Averaged over all $k_{z},$ they
still have the fourfould symmetry and the observable properties should be very
similar to the 2D case. A notable exception is ARPES. That technique should
detect gap nodes along the (0,1) and (1,0) direction when probing $k_{z}=\pi/2c,$
which should gradually shift away from these directions when the probed momentum
is different.

\begin{figure}[t]
\includegraphics[width=0.8\columnwidth]{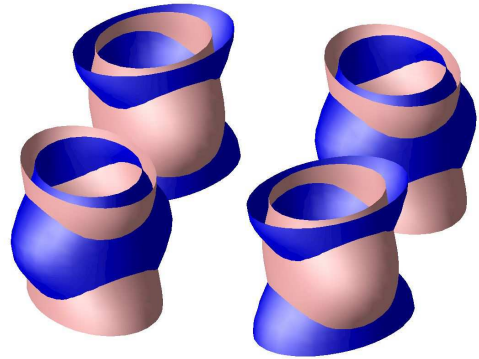}\caption{
Same as Fig. \protect\ref{3D}, but assuming a {\it very strong}
$k_z$ dispersion.  } 
\label{full3D}%
\end{figure}

This is actually the case in density functional calculations for the
stoichiometric compounds in the reported crystal structure; the intersection
lines of the two FSs folded on top of each other never close, and a $d-$wave
superconductivity in this system must retain all four vertical node lines.
Suppose however that these calculations underestimate the $k_{z}$ dispersion
(this is somewhat unlikely, as band structure calculations tend to produce
too diffuse orbitals and too much hopping, but let us
assume for the sake of generality that
this is possible). In that case, at some finite value of $\tilde k_{z}$ such that
$0<\tilde{k}_{z}<\pi/2c$ nodes 1 and 2 will merge and annihilate, and so will
nodes 3 and 4, while at $k_{z}=\pi-\tilde{k}_{z}$ the other two pairs will
annihilate. As a result, we will have a $horizontal$ wiggly node line, the
less wiggly the stronger is the 3D dispersion (Fig. \ref{full3D}. Importantly, a full node line
remains present in any band structure, whatever assumption one makes about the
3D dispersion and ellipticity. \textit{Thus, the fact that fully developed
node lines are inconsistent with numerous reported experiments excludes a
d-wave pairing as a viable possibility.}

\begin{figure}[t]
\includegraphics[width=0.8\columnwidth]{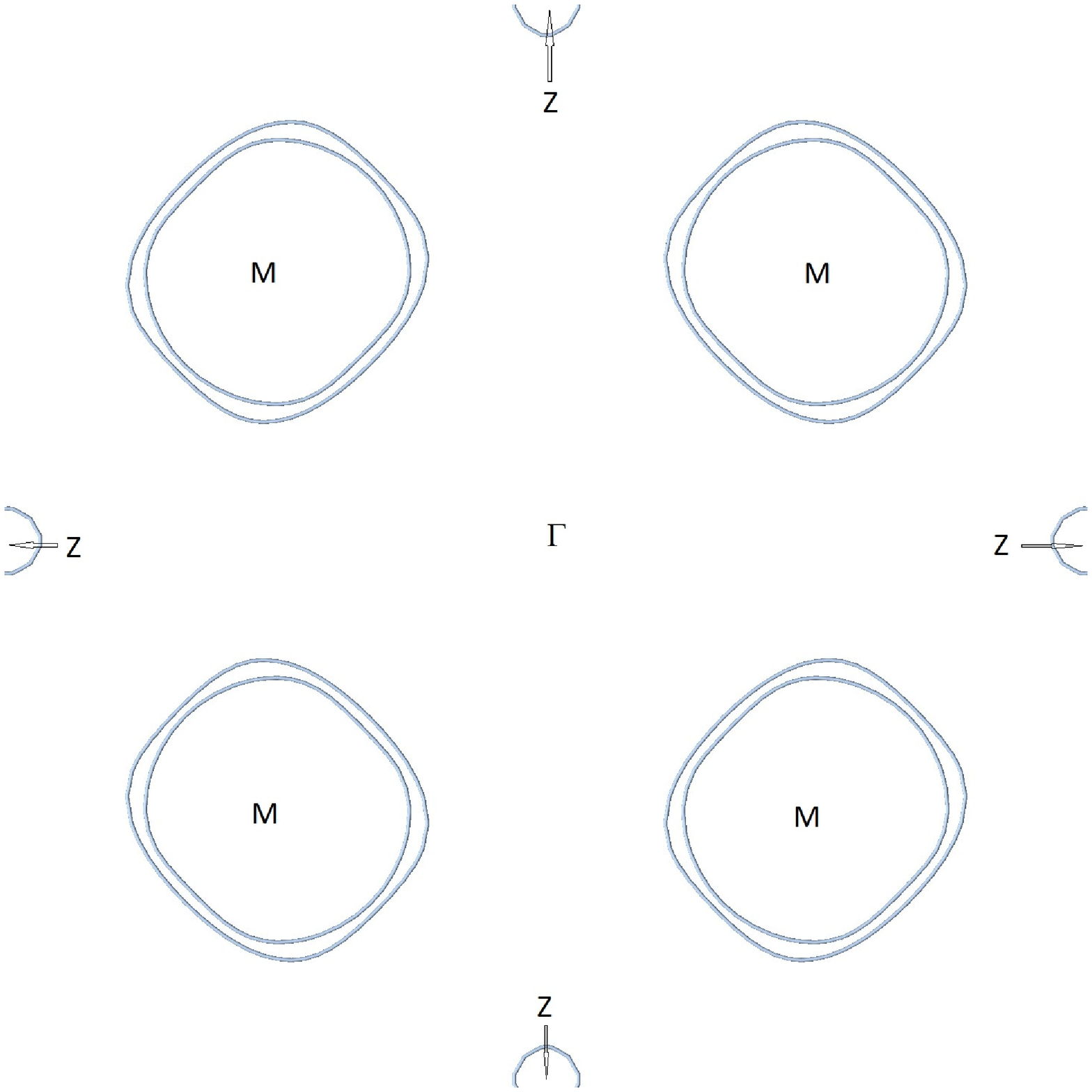}
\includegraphics[width=0.8\columnwidth]{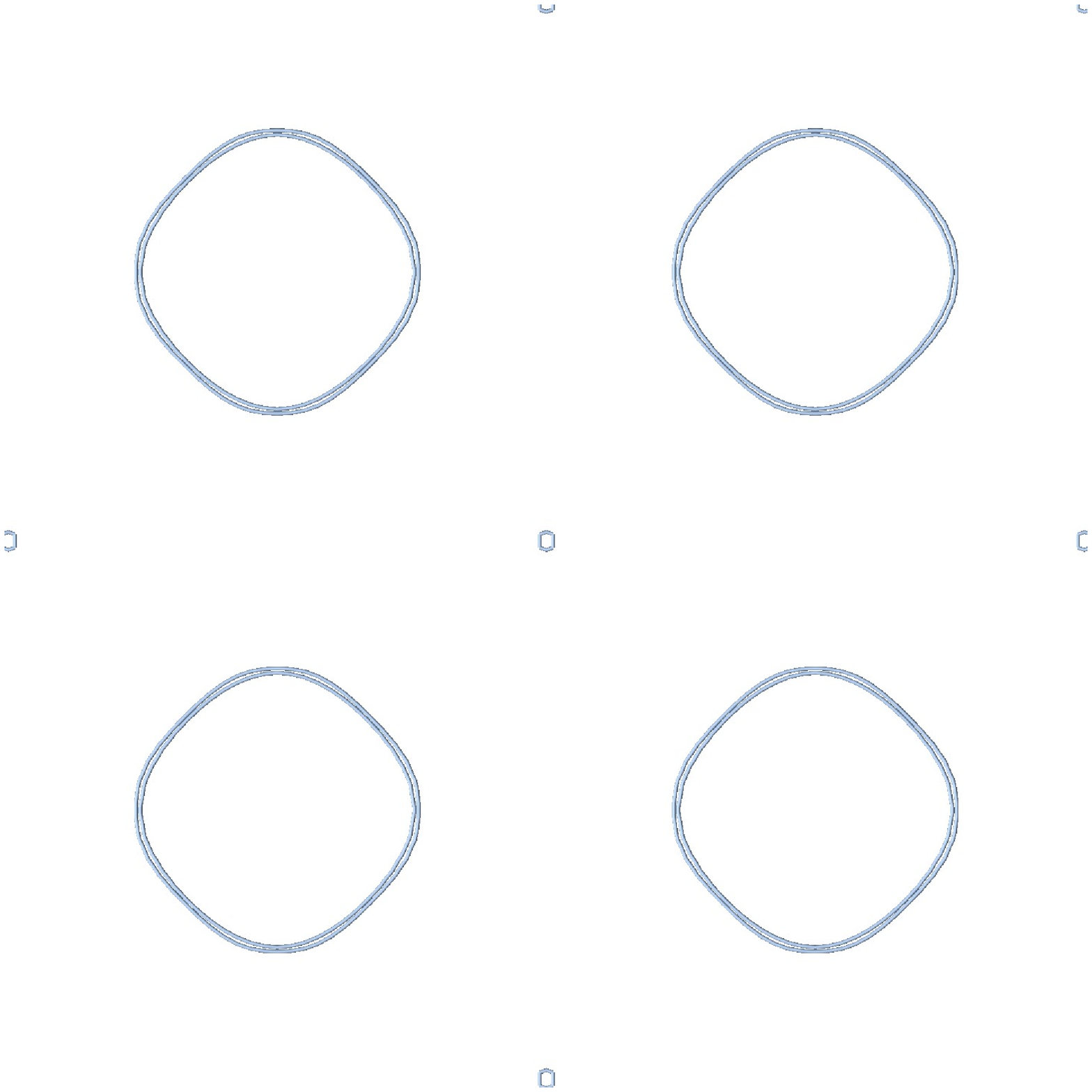}
\caption{
Cuts of the Fermi surfaces calculated for K$_{0.8}$Fe$_2$Se$_2$
using LAPW band structure, and the experimental lattice parameter 
and atomic positions. Upper panel: $k_z=0$. Lower panel:
$k_z=\pi/2c$ (half way between $\Gamma$ and $Z$}.  
\label{GGAFS}%
\end{figure}

An interesting alternative presents itself if we look closely at the
calculated ab-initio Fermi surfaces of KFe$_2$Se$_2$. One feature that distinguishes them from
those in As-based materials is a very small ellipticity and, compared to the
As-based 122 family, very little $k_{z}$ dispersion\cite{notebands}. Looking
at the constant-$k_{z}$ cuts (Fig. \ref{GGAFS}) of the Fermi surface, we observe that we
are in a regime where the separation of the two FSs is comparable with, or
smaller than the hybridization. In this case, a reasonable approximation would
be to neglect both ellipticity and $k_{z}-$dispersion, and analyze the
possible superconducting symmetry in this model. First of all in this
approximation the resulting FSs are two concentric cylinders that touch at
$k_{z}=0$ but are split otherwise. The wave functions on these cylinders are,
respectively, the odd and the even combinations of the original and the
downfolded bands.

Thus, if the pairing interaction in the unfolded BZ exists only in the
interband (interpocket) channel, as is implicitly or explicitly assumed in
most current theories, it becomes identically zero after downfolding and
hybridization. In fact, in this limit, when hybridization is strong everywhere
in the BZ, the spin susceptibility and the pairing interaction must be
computed from scratch using the 2-Fe unit cell (and the folded BZ).

Importantly, one can easily imagine an interaction that would lead to a
\textit{nodeless} state in such a system. Indeed, if the interaction is
stronger between the bonding and antibonding band, than between different
points in the same band, the resulting interaction will again be a
sign-changing s-wave, with all inner barrels having one sign of the order
parameter, and the other the opposite sign (A very
 similar state was unsuccessfully proposed for
bilayer cuprates 15 years ago\cite{bilayer}).

 Naively, one may think that one
can construct a d-wave state where the signs of the order parameter will be
swapped as one goes around from one M point in the BZ to another. Yet this is not
allowed by symmetry, for (2$\pi/a,0,\pi/c)$ and (0,2$\pi/b,\pi/c)$ (2 Fe/cell
notations) are reciprocal lattice vectors, so translating by any of these
vectors must retain both the amplitude and the phase of the superconducting order
parameter. Incidentally, this symmetry requirement is not always appreciated,
and there have been \textquotedblleft$d-$wave\textquotedblright\ suggestions
($e.g.,$ Ref.\cite{Balat}) that violate it.

Let us now discuss possible magnetic interactions in this system.
Both from the Fermiology point of view and from experiment\cite{NMR} it is
clear that familiar spin fluctuations with the wave vector ($\pi/a,\pi
/b,q_{z})$ are absent in this system. As discussed above, model calculations
based on an unfolded band structure are much less well justified than in the
old pnictides, at least if one believes the band structure calculations. In
principle, one can controllably calculate the spin resposne
using the full density functional theory \cite{Serega}, however, there are no codes 
widely available that are implementing such capability. 

On the other hand, one can gain some insight regarding  the DFT spin response at $q=0$, in particular,
on the relative strength of the fluctuations in the FM and in the
AFM (checkerboard) channels, in a different way. To this end, let us write
the full spin suseptibility in the the local density
functional theory\cite{note}:%
\begin{equation}
\chi^{FM}=\frac{\chi_{0}^{FM}}{1-I\chi_{0}^{FM}},\;\chi^{AFM}=\frac{\chi
_{0}^{AFM}}{1-I\chi_{0}^{AFM}},
\end{equation}
where $I=2\delta^{2}E_{xc}/\delta M_{Fe}^{2}$ is the iron Stoner factor, which
we, as the first approximation, will consider independent of the magnetic
pattern. Note that spin-unrestricted calculations for all magnetic patterns,
ferromagnetic, checkerboard, or the stripe phase similar to ferropnictides
converge to large magnetic moment solutions not helpful in analyzing the
linear response of the nonmagnetic phase (Table 1).

\begin{table}
\caption{Calculated energies (the nonmagnetic state
is taken as zero) for various stable and metastable magnetic 
states of KFe$_2$Se$_2$.}
\begin{tabular}{|l|c|c}
\hline
& $M_{Fe},\mu _{B}$ & $\Delta E, $ meV/Fe \\ 
\hline
FM (LDA)&2.8   &$+13$ \\ 
FM (GGA)&2.9  & $-140$ \\ 
AFM-cb (LDA) &1.8  &-111 $$\\
AFM-cb (GGA)& 2.1  &$-192 $\\
stripe (LDA)&2.2&$-169 $\\
stripe (GGA)& 2.4 &$-290$\\
\hline
\end{tabular}
\end{table}

\begin{figure}[t]
\includegraphics[width=0.8\columnwidth]{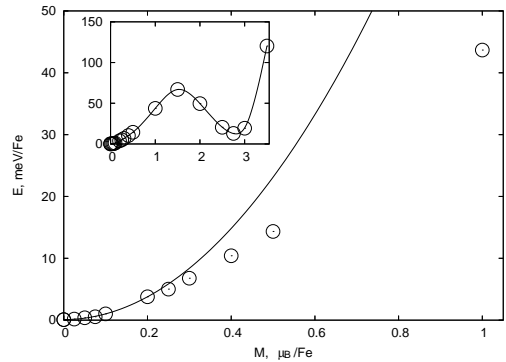}
\caption{Fixed spin moment calculations for the uniform
(ferromagnetic) susceptibility in KFe$_2$Se$_2$.
}  
\label{FSM}%
\end{figure}

To circumvent this problem, we will use a modification of the standard LAPW
package "WIEN2k", which allows for a phenomenological account of itinerant
spin fluctuations by tuning the Hund's rule coupling\cite{Blaha}. It appears
that the unaltered LDA (and even GGA) functional solution in the nonmagnetic phase is stable
against weak FM perturbations (Fig. \ref{FSM}), even though it is unstable
against formation of a large magnetic moment\cite{stripe}. It requires scaling $I$ up by
40\% to make it unstable, thus $\chi_{0}^{FM}\approx1/(1.4I)=0.7I.$ at the
same time, scaling $I$ down by $\alpha\approx0.7,$ we make the checkerboard
pattern also marginally stable, thus $\chi_{0}^{AFM}\approx1/(0.7I)\approx
2\chi_{0}^{FM}.$ Thus, the Fermiology favors the checkerboard
antiferromagnetic fluctuations about twice more than the ferromagnetic ones.

This is in some sense encouraging. If both FM and AFM fluctuations are
present, they can actually provide coupling between the bonding
and antibonding sheets of the folded Fermi surface, even if the hybridization
is very strong (if only AFM fluctuations are present, this coupling vanishes
in the limit of strong hybridization). It may or may not be stronger than the
intraband coupling. Only full calculations of susceptibility in the two Fe
unit cell will give us the answer. Yet, we can firmly conclude that the only
state compatible with two experimental observations, (1) that the
superconducting gap does not have nodes and (2) that superconductivity emerges
in immediate proximity of an ordered magnetic phase, is again an $s_{\pm}$
state, but this time with the order parameter changing sign between the
bonding and antibonding state.
It is also worth noting that if a
3D electron pocket is present at $\Gamma,$ as calculations and several ARPES
experiments suggest, in the proposed $d-$wave symmetry\cite{Graser,Lee,Balat}
it would be cut by four nodal lines which would also have been seen in the
experiment. The concentric $s_{\pm}$ state discussed above does not require
any nodes on this pocket.

Finally, a word of caution is in place. While it is useful, and, arguably,
imperative, at this point of time, to establish the symmetry restrictions on
possible order parameter in AFe$_{2}$Se$_{2}$ compounds, the exprerimental
situation is by far not clear. The compositions reported range from
$\sim$0.8 hole/Fe doped (K$_{0.65}$Fe$_{1.41}$Se$_{2},$ Ref.
\cite{TorchettiNMR}), compared to the stoichiometric AFe$_{2}$Se$_{2},$ to
$\sim 0.9$ electron/Fe (Tl$_{0.63}$K$_{0.37}$Fe$_{1.78}$Se$_{2},$ Ref.
\cite{DingTl}). Se-deficient samples have also been reported\cite{Se-def}.
There have been credible reports about particular ordering of
vacancies\cite{vac}. Yet, the superconducting properties seem to be remarkably
similar. Is it fortuitous that ARPES finds electronic structures remarkably
similar to those computed for stoichiometric compounds, despite large
deviations from stoichometry? More experiments will be needed before we can
gain quantitative understanding. Yet the statements based solely on
crystallographic symmetry, and most of the conclusions of this paper belong to
this class, should hold, and have to be kept in mind.

\acknowledgements
I acknowledge discussions with Andrey Chubukov, Sigfried Graser, 
  Peter Hirschfeld, and Douglas Sclapino. I am partcularly thankful to Ole Andersen and 
Lilia Boeri for helping me figure out the factors that control the ellipticity
and the $k_z-$dispersion in the 122 structure.


\begin{thebibliography}{99}
\bibitem {notebands}This is true for my own calculations, which were performed
in the experimental crystal structure using the standard WIEN2k band structure
package, as well as for other reported calculations\cite{bands1}, although
pseudopotential calculations reprted in Ref. \cite{Graser} have a somewhat
stronger $k_{z}$ dispersion.
\bibitem {note}There are a number of simplifications in this expression, for
instance, we have neglected the so-called local field effects and possible
$\mathbf{q-}$dependence of the Stoner factor $J$, but these simplifications
are not principal for our qualitative estimate.
\bibitem{stripe} Incidentally, the same is true for the ``stripe'' phase, the pattern
that is the ground state in our calculations and is observed experimentally
in pnictides. While the large-moment solution is very stable, the nonmagnetic
state is stable against small perturbations of this symmetry, not only is LDA
but also in GGA.
\bibitem {Blaha}This is implemented by the following recipe: first, at each
iteration, the standard LDA or GGA potential for the spin-up and spin-down
channel is calculated; then, it is rescaled according to formula $v_{\uparrow
}(\mathbf{r})=[v_{\uparrow}(\mathbf{r)+v}_{\downarrow}(\mathbf{r}%
)]/2+\alpha\mathbf{[}v_{\uparrow}(\mathbf{r)-v}_{\downarrow}(\mathbf{r)}%
]/2=v_{\uparrow}(\mathbf{r})(1+\alpha)/2\mathbf{+v}_{\downarrow}%
(\mathbf{r)}(1-\alpha)/2,$ $v_{\downarrow}(\mathbf{r})=v_{\uparrow}%
(\mathbf{r})(1-\alpha)/2\mathbf{+v}_{\downarrow}(\mathbf{r)}(1+\alpha)/2,$
where $0\leq\alpha\leq1.$
\end{thebibliography}
\end{document}